\documentclass{rmaa}
\def\spose#1{\hbox to 0pt{#1\hss}}
\def\lta{\mathrel{\spose{\lower 3pt\hbox{$\mathchar"218$}}
     \raise 2.0pt\hbox{$\mathchar"13C$}}}
\def\gta{\mathrel{\spose{\lower 3pt\hbox{$\mathchar"218$}}
     \raise 2.0pt\hbox{$\mathchar"13E$}}}

\title{Scalings between Physical and their Observationally Related
 Quantities of Merger Remnants}
\author{H\'ector Aceves\altaffilmark{1} \& H\'ector Vel\'azquez
  \affil{Instituto de Astronom\'{\i}a, UNAM, Ensenada}
  }
\altaffiltext{1}{Email: \texttt{aceves@astrosen.unam.mx}}
\fulladdresses{
\item H\'ector Aceves \& H\'ector Vel\'azquez:
  Instituto de Astronom\'{\i}a, UNAM Campus Ensenada,
  Apdo Postal 877, Ensenada 22800, Baja California, M\'exico
  (aceves@astrosen.unam.mx, hmv@astrosen.unam.mx)}

\shortauthor{Aceves \& Vel\'azquez} \shorttitle{Scalings in Mergers}
\SetVolume{00} \SetFirstPage{000} \SetYear{2004}
\ReceivedDate{------}
\AcceptedDate{------}
\SetYear{2005}

\resumen{Se presentan relaciones de escala entre la velocidad 
virial ($V$)
y la dispersi\'on de velocidades uni-dimensional central ($\sigma_0$); 
el radio gravitacional
 ($R_{\rm v}$) y el radio efectivo ($R_{\rm e}$); y la masa total ($M$) y la
masa luminosa ($M_{\rm L}$) encontradas en simulaciones de $N$-cuerpos de
fusiones binarias entre galaxias espirales. 
Estas relaciones son de la forma $V^2 \propto \sigma_0^\alpha$, 
$R_{\rm v} \propto R_{\rm e}^\beta$ y $M\propto M_{\rm L}^\gamma$.
Los valores particulares obtenidos para  $\{\alpha,\beta, \gamma\}$
dependen del m\'etodo de ajuste utilizado [m\'{\i}nimos cuadrados ordinarios
 ({\sc ols}) o regresi\'on ortogonal de distancia ({\sc odr})], 
el tipo de perfil supuesto [de Vaucouleurs (deV) or S\'ersic (S)], y el
tama\~no del intervalo radial donde se hace el ajuste.
Los \'{\i}ndices $\alpha$ y $\gamma$ resultan ser los m\'as sensibles al
procedimiento de ajuste, obteni\'endose
para el {\sc ols} un promedio $\langle \alpha \rangle_{\rm ols}=1.51$  y  
$\langle \gamma \rangle_{\rm ols}=0.69$,  mientras que para el  {\sc odr} 
un  $\langle \alpha \rangle_{\rm odr}=2.35$ y 
 $\langle \gamma \rangle_{\rm odr}=0.76$. 
El \'{\i}ndice $\beta$ depende m\'as de el perfil adoptado, con
$\langle \beta \rangle_{\rm{deV}}=0.13$ y 
$\langle \beta \rangle_{\rm{S}}=0.27$.
Conclu\'{\i}mos que los remanentes de fusiones formados de manera no disipativa
tienen un rompimiento fuerte de homolog\'{\i}a estructural y cinem\'atica.
}

\abstract{We present 
scaling relations between the virial velocity ($V$) and 
the one-dimensional central
velocity dispersion ($\sigma_0$); the gravitational radius 
($R_{\rm v}$) and the
effective radius ($R_{\rm e}$); and the total mass ($M$) and the luminous 
mass ($M_{\rm L}$) 
found in $N$-body simulations of binary mergers of spiral galaxies.
These scalings are of the form $V^2 \propto \sigma_0^\alpha$, 
$R_{\rm v} \propto R_{\rm e}^\beta$ and $M\propto M_{\rm L}^\gamma$.
The particlar values obtained for $\{\alpha,\beta, \gamma\}$ depend on the
method of fitting used [ordinary least-squares ({\sc ols}) or orthogonal 
distance regression ({\sc odr})], 
the assumed profile [de Vaucouleurs (deV) or S\'ersic (S)], and the size
of the radial interval  where the fit is done. 
The $\alpha$ and 
$\gamma$ indexes turn out more sensitive to the fitting procedure, obtaining
for the {\sc ols} a mean $\langle \alpha \rangle_{\rm ols}=1.51$  and 
$\langle \gamma \rangle_{\rm ols}=0.69$,  while for  the {\sc odr} 
  $\langle \alpha \rangle_{\rm odr}=2.35$ and  
$\langle \gamma \rangle_{\rm odr}=0.76$. 
The $\beta$ index depends more on the adopted type of profile, with 
$\langle \beta \rangle_{\rm{deV}}=0.13$ and 
$\langle \beta \rangle_{\rm{S}}=0.27$.
We conclude that dissipationless formed remnants of mergers have 
 a strong breaking of structural and kinematical 
homology.
}

\begin{document}
\maketitle

\section{Introduction}\label{sec:intro}

Toomre's (1977) idea that the merging of spirals could lead to an
elliptical galaxy has found ground evidence, both theoretical and
observational (e.g., Barnes 1998, Schweizer 1998), although some questions
remain open (e.g., Peebles 2002, Chiosi \& Carraro 2002).

Ellipticals show a number of regularities among their kinematical and
structural properties that have been recognized in the past, such
as the Kormendy and Faber-Jackson relation, and the Fundamental
Plane of Ellipticals (e.g., Kormendy 1977, Faber \& Jackson 1976, 
Bernardi~et~al.~2003a,b).

Understanding the physical origin of these relations is important since they 
are intimately related to their formation and evolutionary history.
Obtaining theoretical scalings relations among physical and
observational quantities depends on our knowledge, for example, 
of the  star formation process, the  distribution of dark matter in
ellipticals, and the kinematics of remnants of spiral mergers.
Given the complexities of the problem, we restrict ourselves
here to obtain scaling relations resulting solely from dissipationless
simulations of mergers of spirals. 

 In a related paper (Aceves~\&~Vel\'azquez~2005) the accumulated effects of 
spatial and kinematical homology breaking on the determination of a Fundamental
Plane (FP) like relation for remnants was studied, but no detailed examination
of how the different physical quantities involved scaled with their 
observational counterparts. In this work we address this matter in more detail,
and include two more merger simulations than in Aceves~\& Vel\'azquez (2005).

In particular, we consider here only  quantities
involved in the virial theorem and we determine their 
dependences with their observational counterparts:
\begin{equation}\label{eq:relations}
V^2 \propto \sigma_0^\alpha, \quad
R_{\rm g} \propto R_{\rm e}^\beta, \quad
M \propto M_{\rm L}^\gamma \,;
\end{equation}
where $V$ is virial velocity, $\sigma_0$ the one-dimensional central 
velocity dispersion, $R_{\rm g}$ the gravitational radius,
$R_{\rm e}$ the effective radius (i.e., that enclosing half of the 
luminous matter), $M$ the total mass of the system,
and $M_{\rm L}$ the luminous mass. 
Homology between the physical and observational quantities requires that
$\alpha=2$, and $\beta=\gamma=1$ in expression (\ref{eq:relations}).

A further motivation for this study stems from the fact that behavior of
the previous relations bears direct
impact on the estimate of a dynamical mass $M_{\rm dyn}$
 in ellipticals (e.~g., Padmanabhan et~al.~2004). It is common to assume
 in such studies 
an homologous relation of the form 
$M_{\rm dyn} \propto \sigma_0^2 R_{\rm e}$. 
However, as we will see, merger remnants show an important deviation 
from homology that most probably reflect the actual situation in ellipticals.
Thus this effect of non-homology  would need to  be taken into consideration
  when estimating the dynamical mass of ellipticals. However,
 the study of such a 
a problem  is out of the scope of the present work.

This paper has been organized as follows. In Section~2, we summarize the
numerical models of spiral galaxies used.  We describe the initial conditions 
for the
encounters as well as we provide some information on the computational aspects
on the simulations carried out. 
 In Section~3 we present the method used to determine the
physical quantities $\{V,R_{\rm g},M\}$ and their observationally related
ones  $\{\sigma_0, R_{\rm e}, M_{\rm L}\}$. Also, the fitting procedures 
used to obtain the scaling indexes
 in  (\ref{eq:relations}) are indicated.
In Section~4, the results obtained for $\{\alpha,\beta, \gamma\}$
in (\ref{eq:relations}) are presented. Finally, in Section~5, we 
summarize our main conclusions.

\section{Galaxy models and initial conditions for encounters}

\subsection{Galaxy Models}

Our spiral galaxy models consist of a spherical dark halo and a stellar disk 
component. The contribution of a central bulge is not considered here. 
The disk profile has the functional form
\begin{equation}
\rho_{\rm d} (R,z) = \frac{M_{\rm d}}{4 \pi R^2_{\rm d} z_d} \exp( -
R/R_{\rm d})  \, {\rm sech}^{2} (z/z_d) \;,
\end{equation}
where $R_{\rm d}$ and $z_{\rm d}$ are the radial and vertical
scale-lengths
of the disk, respectively.
The vertical length, $z_{\rm d}$, is randomly taken from the interval $(0.1-0.2)R_{\rm d}$;
where $R_{\rm d}$ is obtained as indicated below.

The dark halo follows a Navarro, Frenk \& White (NFW, 1997) profile,
 modified with an exponential cutoff:
\begin{equation}\label{eq:nfwm}
\rho_{\rm h} (r) = \frac{M_{\rm h}\, \alpha_{\rm h} }{4 \pi r  (r +
r_{\rm
s})^2}
\ \exp\left[ - \left(\frac{r}{r_{200}} + q \right)^2  \right]  \;,
\end{equation}
with
$$
\alpha_{\rm h} = \frac{\exp( q^2) }{ \sqrt{\pi} q \exp(q^2) {\rm
Erfc}(q) +
\frac{1}{2} \exp(q^2) {\rm E}_1(q^2) - 1 } ;
$$
 where ${\rm Erfc}(x)$ is the
complimentary error function and ${\rm E}_1(x)$
the exponential integral.
The scale radius of the dark matter profile is $r_{\rm s}$, 
$c=1/q=r_{200}/r_{\rm s}$ is the concentration, and $M_{\rm h}$ is the halo mass; 
 $r_{200}$ is defined as the radius where the mean interior density 
is 200 times the critical density.

The properties of the disk are set up satisfying the Tully-Fisher relation
(Tully \& Fisher 1977, Giovanelly et~al.~1997). This is carried out by 
following
the study of  Shen, Mo \& Shu
(2002, hereafter SMS) and using the disk galaxy formation model of Mo, Mao
\& White (1998, hereafter MMW); from which we can obtain $R_{\rm
d}$.

In the MMW framework five parameters are required to obtain the radial
scale-length of the disk. These are the circular velocity $V_{\rm c}$
at
$r_{200}$, the dimensionless spin parameter $\lambda$,  the
concentration $c$ of the dark halo, the fraction of disk to
halo mass $m_{\rm d}$,  the fraction of angular momentum
in the disk to that in the halo $j_{\rm d}$. We
have followed the procedure outlined by SMS
 to construct our galaxy models, and chosen an epoch for the formation
of disks at a redshift of $z=1$ (Peebles 1993).

We have selected only spirals with circular
velocities in the range from 50 to 300 km~s$^{-1}$, and with a  
disk stability
parameter $\varepsilon_{\rm m}$$\,\ge\,$$0.9$; where
$\varepsilon_{\rm m}$=$V_{\rm m} (G M_{\rm d} / R_{\rm d})^{-1/2}$ 
and $V_{\rm m}$ is the maximum rotation velocity (Efsthatiou, Lake
\& Negroponte 1982, Syer, Mao \& Mo 1997). From an ensemble of random
points
constructed according to the scheme of SMS, and satisfying the previous
conditions, we obtained the final properties of the 
24 galaxies that take part in our 12 binary 
merger simulations. 

In Table~\ref{tab:glxmodels} the particular 
values for each galaxy model constructed are listed; 
where $N_{\rm h}$ and $N_{\rm d}$
are the number of particles used in the halo and disk, respectively.
The last column lists the pericenter 
radius $R_{\rm p}$ for the encounters, assuming that galaxies are point
particles.
Finally, Hernquist's method (1993) was used to set up the particle 
velocities in our self-consistent models.

\begin{table*}[!t]
 \begin{minipage}{140mm}
  \caption{Properties of Initial Galaxies}\label{tab:glxmodels}
  \begin{tabular}{l|crccrr|cccr|r}
  \hline
   Merger     &  \multicolumn{6}{c}{Halo} & \multicolumn{4}{c}{Disk} &\\
    & $M_{\rm h}$ & $r_{200}$ & $\lambda$ & $j$
 & $c$ & $N_{\rm h}$ & $M_{\rm d}$ & $R_{\rm d}$ & $z_{\rm d}$ & $N_{\rm d}$ & $R_{\rm p}$ \\
    & [M$_\odot$] & [kpc] &  & &
     &  & [M$_{\odot}$] & [kpc] & [kpc] &  & [kpc] \\
 \hline
 $M01$  & $4.02\times 10^{11}$ & 104.3 & 0.063 & 0.041 & 7.63 & 57126 & $2.11\times   10^{10}$ & 2.4 & 0.39 & 12000 & 13.2\\
      & $1.25\times 10^{12}$ & 152.3 & 0.056 & 0.068 & 3.84 & 177571 & $7.78\times 10^{10}$ & 5.7 & 0.47 & 44267 & \\
\hline
 $M02$  & $1.39\times 10^{11}$ & 73.3 & 0.021 & 0.052 & 8.94 & 172403 & $4.48\times 10^{9}$ & 1.2 & 0.12 & 57436 & 13.3 \\
      & $6.45\times 10^{10}$ & 56.7 & 0.028 & 0.033 & 6.61 & 80000 & $1.56\times 10^{9}$ & 1.3 & 0.13 & 20000 & \\
\hline
 $M03$  & $8.65\times 10^{10}$ & 62.6 & 0.078 & 0.033 & 11.12 & 149138 & $3.62\times 10^{9}$ & 1.7 & 0.20 & 42093 & 6.9  \\
      & $4.64\times 10^{10}$ & 50.8 & 0.031 & 0.043 & 12.38 & 80000 & $1.72\times   10^{9}$ & 0.7 & 0.12 & 20000 & \\
\hline
 $M04$  & $2.36\times 10^{11}$ & 87.4 & 0.036 & 0.029 & 9.17 & 80000 & $8.64\times 10^{9}$ & 1.1 & 0.19 & 20000 & 10.4 \\
      & $1.75\times 10^{11}$ & 79.2 & 0.048 & 0.054 & 5.29 & 59428 & $6.77\times 10^{9}$ & 3.3 & 0.58 & 15671 & \\
\hline
 $M05$  & $5.42\times 10^{10}$ & 53.5 & 0.053 & 0.114 & 15.95 & 74829 & $4.82\times 10^{9}$ & 1.2 & 0.19 & 20000 & 18.6  \\
      & $6.02\times 10^{10}$ & 55.4 & 0.034 & 0.077 & 12.05 & 83124 & $3.44\times 10^{9}$ & 1.0 & 0.14 & 14245 & \\
\hline
 $M06$  & $7.49\times 10^{10}$ & 59.6 & 0.098 & 0.059 & 7.91 &91546 & $5.11\times 10^{9}$ & 2.5 & 0.35 & 25000 & 15.5 \\
      & $6.68\times 10^{10}$ & 57.4 & 0.078 & 0.043 & 13.32 & 81742 & $2.93\times 10^{9}$ & 1.9 & 0.23 & 14327 & \\
\hline
 $M07$  & $4.88\times 10^{10}$ & 51.7 & 0.074 & 0.064 & 11.02 & 73643 & $3.31\times 10^{9}$ & 1.5 & 0.25 & 20000 & 14.1 \\
      & $9.48\times 10^{10}$ & 64.5 & 0.047 & 0.103 & 11.79 & 143062 & $6.98\times 10^{9}$ & 1.6 & 0.31 & 42119 & \\
\hline
 $M08$  & $9.77\times 10^{10}$ & 65.2 & 0.032 & 0.034 & 6.56 & 245462  & $2.39\times 10^{9}$ & 1.7 & 0.20 & 36000 & 12.5 \\
      & $1.02\times 10^{11}$ & 66.1 & 0.023 & 0.012 & 7.80 & 188124 & $1.31\times 10^{9}$ & 0.9 & 0.12 & 19725 & \\
\hline
 $M09$ & $8.11\times 10^{10}$ & 61.2 & 0.099 & 0.142 & 10.87 & 150000 & $7.22\times 10^{9}$ & 4.6 & 0.73 & 30000 & 8.4 \\
  & $8.33\times 10^{10}$ & 61.8 & 0.122 & 0.095 & 10.01 & 154050 & $7.15\times 10^{9}$ & 3.9 & 0.56 & 29686 & \\
\hline
$M10$ & $1.27\times 10^{11}$ &  $71.0$ & $0.109$ & $0.078$ & $ 9.42$ & 240000 &
$ 9.14\times 10^{9}$ & $4.0$ & $0.42$ & 60000 & 7.9 \\
 &  $8.76\times 10^{10}$ & $ 62.8$ &  $0.088$ & $0.089$ & $9.23$ & 165992 &
 $7.42 \times 10^{9}$ & $2.9$ &  $0.46$ & 48689 & \\
\hline
$M11$ &  $4.74\times 10^{11} $ & 110.3 & 0.110 & 0.106 & 11.4 & 542682 & $4.16 \times 10^{10}$ & 6.6 & 1.29 & 289205  & 5.2 \\
 & $6.99\times 10^{10}$ &  58.3  & 0.071  & 0.035  &   9.94 & 80000 & $2.88\times 10^9$  &  1.6  & 0.19 & 20000 & \\
\hline
$M12$ & $5.65\times 10^{10}$ &  54.3 &  0.076 & 0.103  &  7.21 & 60000 & $4.65\times 10^9$ &  2.6 &  0.35 & 20000 & 17.7\\
  & $3.11\times 10^{11}$ &  95.9  & 0.056 & 0.075  &  7.41 & 330909 &  $1.96 \times 10^{10}$  & 3.1  & 0.46 & 84596 & \\
\hline
\end{tabular}
\end{minipage}
\end{table*}

\subsection{Encounter Parameters}

A large number of simulations, and computational resources, would
be required to sample the parameter space of binary encounters of disk
galaxies in order to address  the  dynamical effects of, 
for example, the 
pericenter distance and disk orientations on the scaling indexes $\{\alpha,
\beta,\gamma\}$ in (\ref{eq:relations}). 

We decided instead to sample randomly the encounters 
 initial conditions, but considering only parabolic encounters. 
The pericenters $R_{\rm p}$ were chosen randomly
in the range of $\{5-20\}$~kpc; values that are typically found in 
cosmological
simulations and that tend to favor mergers (e.g., Navarro, Frenk \& White
1995). The particular values of $R_{\rm p}$ are indicated in  
Table~\ref{tab:glxmodels}. 

The initial separation between two galaxies is 25\% larger than
the sum of their corresponding $r_{200}$ radii. The spin orientation of
each galaxy, relative to the orbital plane, is taken also randomly.

\subsection{Computational Issues}

The simulations were done using {\sc
  GADGET}, a tree-based code (Springel, Yoshida \& White 2001), and run
on  a Pentium cluster of 32 processors (Vel\'azquez \&
Aguilar  2003). We chose the softening parameter for disk particles
$\epsilon_{\rm d} \!=\! 35\,$pc and $\epsilon_{\rm h}\!=\!350\,$pc for
dark particles.

{\sc GADGET} uses a spline kernel for the softening, so
the gravitational interaction between two particles is fully Newtonian
for separations larger than twice the softening parameter
(Power et al. 2003).  This corresponds in practice to the numerical
resolution of our simulations.

We evolved in isolation the
numerical realizations of each galaxy for about $2\,$Gyr, and no
significant change was appreciated in their density profiles or virial ratio.
Each binary merger was followed for a total
time of about $8\,$Gyr.
 At this time the remnants had reached a stable virial ratio.
The typical time of arrival to pericenter is
about $1\,$Gyr.  
Energy conservation was better than 0.25\% in 
all simulations. 
Each simulation took $\approx 2$ weeks of wall clock time in our PC cluster.

The center of each remnant was determined by the center-of-mass of the
1\% most bounded particles.
We eliminated any residual bulk motion from the remnant before computing
their properties.

\section{Method}

In this section we describe how the different physical and observational
 quantities were obtained. Also,
the different fitting procedures used to obtain $\{\alpha,\beta,\gamma\}$ in 
(\ref{eq:relations}) are indicated.

\subsection{Physical Quantities}

The virial relation may be written as $V^2 = G M/R_{\rm g}$;
where $V^2$ is the 3-dimensional velocity dispersion, $M$ the total
mass of the system and $R_{\rm g}$ the gravitational radius. We estimate
these quantities as: $V^2 = 2 T/M$ and $R_{\rm g} = G M^2/|W|$, 
where $T$ and $W$ are
the kinetic and gravitational energy of the remnants, respectively.

 The total kinetic
energy $T$ and gravitational energy $W$ were computed from the usual formulae:
\begin{equation}
T = \frac{1}{2} \sum_{i=1} m_i v_i^2 \;, \quad
W = - G \sum_{i}\sum_{i<j} \frac{m_i m_j}{ r_{ij} } \;,
\end{equation}
where $r_{ij}$ the separation between particles $i$-th and $j$-th. 
 The summation is taken only over the bound particles of the resulting remnant.

\subsection{Observational Quantities}

In order to obtain $R_{\rm e}$ an observational ``procedure''  was followed.
 We fitted an assumed profile to the
surface density profile of the luminous mass, $\Sigma (R)$. We choose
for this the $R^{1/4}$-profile (de Vaucouleurs 1948) and S\'ersic
 $R^{1/n}$-profile (S\'ersic 1968). These are analytical formulae
 that are commonly used in observational studies of 
ellipticals (e.g., Caon, Capaccioli \& D'Onofrio 1993).

 S\'ersic profile has the form
\begin{equation}\label{eq:profile}
\Sigma (R) = \Sigma_0 \exp[ -b (R/R_{\rm e})^{1/n}] \;;
\end{equation}
where $b=b(n)$. This profile reduces to de Vaucouleurs one when the index
$n=4$; for $n=1$ an exponential profile is obtained.

Another observational procedure to determine the structural parameters 
is by fitting the growth-curve of the luminous component
 (e.g., Burstein et~al.~1987, Prugniel \& Simien~1997, Binggeli \&
 Jerjen 1998).

The accumulated luminous mass (or growth-curve) for a S\'ersic law is 
\begin{eqnarray}\label{eq:curve}
M_{\rm L}(R) &=& 2 \pi \int^R \Sigma(R) R \, {\rm d}R \nonumber \\
& =& \frac{ 2 \pi n }{b^{2n}} \, \Sigma_0 R_{\rm e}^2 \,
\gamma(2n,b x^{1/n})
\end{eqnarray}
where $x=R/R_{\rm e}$, and (for $\alpha >0$)
$$
\gamma(\alpha,x) = \int_0^x {\rm e}^{-t} t^{\alpha-1}\, {\rm d}t
$$
is the incomplete gamma function (Ciotti \& Bertin 1999).
The total luminous mass is given by
\begin{equation}\label{eq:mltotal}
M_{\rm L} = \frac{2 \pi n }{b^{2n} }  \Gamma(2n)\, \Sigma_0 R_{\rm e}^2
\,,
\end{equation}
were $\Gamma$ is the complete gamma function.  We assume here the
approximation $b(n)=2n-1/3+4/(405 n)+ 46/(25515n^2)$, that provides a
relative error of $\lta 10^{-7}$ in the range $n\in(1,10)$.

The structural parameters are somewhat dependent
on whether a profile or growth-curve is fitted. Hence, we 
 have fitted both a density
profile, $\Sigma(R)$, and a growth-curve, $M_{\rm L}(R)$,
 to the luminous component of our remnants. We obtained the parameters
$R_{\rm e}$ and $M_{\rm L}$ by  minimizing the $\chi^2$ using 
the Levenberg-Marquardt method (Press et al. 1992).

The central velocity dispersion of all the luminous particles was computed
inside a circular region of projected radius $R_{\rm e}/8$; a standard practice
in observational studies (e.g., J{\o}rgensen, Franx, Kjaergaard 1996).
Thus, the particular value of $\sigma_0$ depends on the 
$R_{\rm e}$ obtained from the adopted method of fitting.
 We computed $\sigma_0$ as
\begin{equation}
\sigma_0 = \sqrt{\frac{1}{N-1} \sum_{i=1}^{N} ( V_{z_i} -
\langle V_z \rangle )^2   }  \;;
\end{equation}
where $N$ corresponds to the number of particles inside the aperture, 
$V_{z_i}$ 
is the line-of-sight velocity of the $i$-th particle, and 
$\langle V_z \rangle$ is the mean velocity integrated along the line-of-sight.

In order to have a better statistics and mimic observations,
 we looked at each remnant along 100 random
different lines-of-sight. For each projection we computed $R_{\rm e}$, 
$M_{\rm L}$ and $\sigma_0$ as described above. 
This conforms our data set over which the linear fits are used to
determine the scaling indexes.

\begin{figure}[!t]
\centering
\includegraphics[width=8cm]{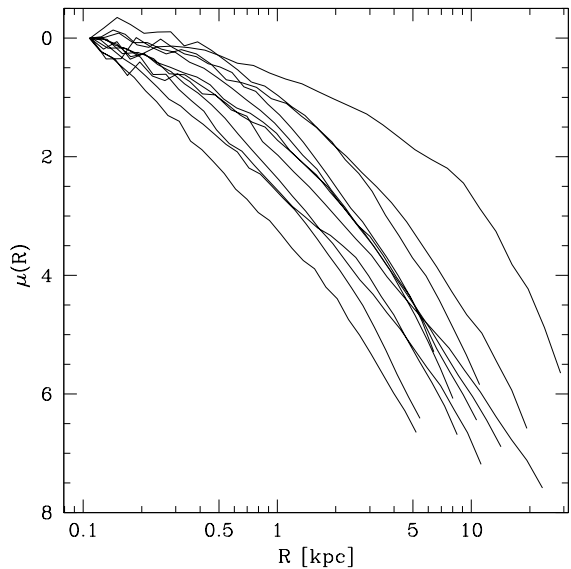}
\vspace{-0.8cm}
\caption{Surface density profiles in ``magnitudes'' for all of our merger 
remnants; normalized to their ``central'' value. 
The outer limit of the fitting interval is taken to be 
$3R_{\rm l}$, where $R_{\rm l}$ is the actual half-luminous mass radius.
}
\label{fig:perfiles}
\end{figure}

\subsection{Interval of Fitting}

The structural parameters depend on 
 the region where the fit of the light profile is done
 (e.g., Caon~et~al.~1993;
Graham 1998; Bertin, Ciotti \&  Del Principe 2002). 
Observationally, the significant radial range for the fit
goes, for example, beyond the region dominated by seeing effects 
to that where the data is considered reliable.

We have considered two inner radii $\xi$ for the fitting interval.
First,  we take a value of  $\xi\!=\!100\,\textrm{pc}\approx
3 \epsilon_{\rm d}$; the numerical resolution of our simulations
for the luminous component. Secondly, we 
recalculated the scalings now adopting a value of
 $\xi\!=\! 1\,$kpc. As a reference, 
in a flat universe 
($\Omega_{\rm m}=0.3$, $\Omega_\Lambda=0.7$) with Hubble's parameter
 $h=0.7$ an
angular size of $1''.5$ at, for example, the Coma Cluster ($z=0.023$)
corresponds to a physical size of $\approx 700\,$pc.

Contrary to observations, in $N$-body simulations we can sample the complete 
luminous component and hence  determine accurately the
 half-light radius $R_{\rm l}$ of the remnant; 
that would correspond under ideal
circumstances to the definition of $R_{\rm e}$.  
However, given that $R_{\rm e}$ is obtained from a fitting procedure, its 
value
will be dependent on the radial range covered by the fitting. 
In order to minimize such bias,
we have done fittings inside a circular aperture of 
radius $3R_{\rm l}$. At this outer radius the range of sampled ``magnitudes''
 [$\mu=-2.5 \log \Sigma(R)$] of the luminous component
is in average $\approx \!7$; see Figure~\ref{fig:perfiles}. This interval
 in  magnitudes is similar to the found in some observational studies 
of ellipticals  (e.g., Prugniel \& Simien 1997, Bernardi et~al.~2003a).

\subsection{Fitting Procedure}

To compute $\{\alpha,\beta, \gamma\}$ in  the expressions of
 (\ref{eq:relations})
a least-squares linear fit, in log-space,  was performed. Different
methods exist to do such fit, and it is known that the slope of the
fitted line depend on the fitting procedure (e.g., Feigelson \& Babu 1992). 

Here the ordinary least-squares ({\sc ols}) and the orthogonal
distance regression ({\sc odr}) procedures are used. The {\sc odr} method
is particularly useful  when there is no clear distinction between the dependent or
independent variable, and when both variables contain 
uncertainties. Moreover, the {\sc odr} fit is insensitive to whether 
the data are weighted or not (Wu, Fang, Xu 1998).
The {\sc odr} line fitting is described  in 
Appendix A.

In order to estimate the standard deviations in the 
scaling indexes $\{\alpha,\beta,\gamma\}$ a
 bootstrap technique (Efron \& Tibshirani 1993)
 was used on our data set.

\section{Scaling Relations}

In this section, we present the scaling relations, among the physical 
quantities that appear in the virial relation and their corresponding
observational counterparts, obtained from our merger remnants.

\begin{figure}[!t]
\centering
\includegraphics[width=8cm]{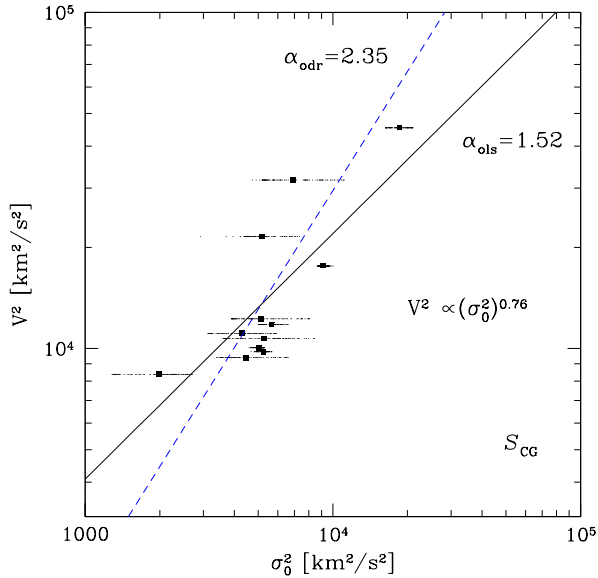}
\vspace{-0.8cm}
\caption{Virial velocity against the central velocity
 dispersion for each projection ({\it dots})
 of a remnant; average values are indicated by  solid squares. 
The solid line
 corresponds to an {\sc ols} fit to the data obtained using a S\'ersic
 growth-curve technique and $\xi\!=\!100\,$pc; the scaling expression 
for this case is indicated. The broken line represents the fit by {\sc odr}.
 The slope of each line is indicated.}
\label{fig:vels}
\end{figure}

\subsection{Velocities}

In Figure~\ref{fig:vels}
 we plot, in log-space, $V^2$ against $\sigma^2_0$ for each
projection ({\it dots}) of the remnants, where the average value of
$\sigma_0^2$ is indicated by a larger symbol.
An homologous behavior for the velocity scaling would require 
that $V^2 \propto \sigma_0^2$.

\begin{table}
\begin{center}
\centering
\caption{Velocity scalings $V^2\propto\sigma_0^\alpha$}
\begin{tabular}{|r|c|c|c|c|c|} \hline
$\xi$  & $R^{1/4}_{\rm P}$ & $R^{1/4}_{\rm{CG}}$ & $S_{\rm P}$ & $S_{\rm{CG}}$
 & {\sc FIT}    \\ \hline \hline
100 pc &  1.53 & 1.58 & 1.52 & 1.52 & {\sc ols} \\
       &  2.40 & 2.40 & 2.36 & 2.35 & {\sc odr} \\ \hline
1000 pc & 1.46 & 1.54 & 1.48 & 1.46 & {\sc ols} \\
        & 2.30 & 2.40 & 2.33 & 2.30 & {\sc odr}  \\  \hline
\end{tabular}
\end{center}
{\footnotesize
$R^{1/4}$ and $S$ indicate a de Vaucouleurs and S\'ersic profile, respectively.
 Subscripts $\rm{P}$ and $\rm{GC}$ refer to whether a fitting  to the surface 
 density profile $\Sigma(R)$ or the curve of growth $M_{\rm L}(R)$ is done, 
and {\sc ols} and {\sc odr}
  refer to the type of least-squares fitting procedure used. }
\label{tab:velocity}
\end{table}

A  relation of the form
\begin{equation}
V^2  \propto \sigma_0^\alpha \,,
\end{equation}
in log-space, was done to out data set.
 In Table~2 we list the results 
obtained for the $\alpha$ index under both fitting procedures {\sc odr}
and {\sc ols} used, and both 
inner boundary radii $\xi$.  
The error in   $\alpha$  by the {\sc ols} procedure is $\pm 0.01$, 
while that for the {\sc odr} procedure  is $\pm 0.02$.

It can be noted that both fitting procedures ({\sc ols} and {\sc odr})
lead to different values for the $\alpha$ index. 
The {\sc odr} procedure
gives a value of $\alpha$ about 50\% higher than that
obtained by {\sc ols}, under all the conditions used for the fittings.

Increasing $\xi$ in the fit leads to a small 
change ($\lta~5$\%) in the $\alpha$ index for the same fitting procedure.
 At a fixed
 $\xi$,  using the different fitting functions (profile or growth-curve),  
  results in small variations ($< 5$\%) in the value of $\alpha$.

Averaging the results in Table~2, over  $\xi$ and type of 
profile considered,
we have for the  {\sc ols} procedure a $\langle \alpha \rangle_{\rm{ols}}
 = 1.51$  while  the {\sc odr} approaches a value of 
 $\langle \alpha \rangle_{\rm{odr}} = 2.35$. 
In general, the values of $\alpha$  in our remnants
 deviate by $\approx 20$\% from the homology expected value of $\alpha=2$.
It follows from these results that our
merger remnants do not satisfy kinematical homology.

We recall that 
the total (random plus rotational) kinetic energy of the system is related
to $V^2$, and $\sigma_0^2$ is essentially a measure of random motion.
The rather small departure from homology of  $\alpha$ seems to suggest
that the contribution from rotational energy might be small
 in the luminous part of the remnants. This result appears 
consistent with some observations of the rotational contribution to kinetic
energy in ellipticals (e.g., Prugniel \& Simien 1994).

However, to test the degree of rotational energy in our remnants a detailed
kinematical analysis would be required, a topic that is under
investigation at this time and to be presented in a future work.

\subsection{Radii}

\begin{figure}[!t]
\centering
\includegraphics[width=8cm]{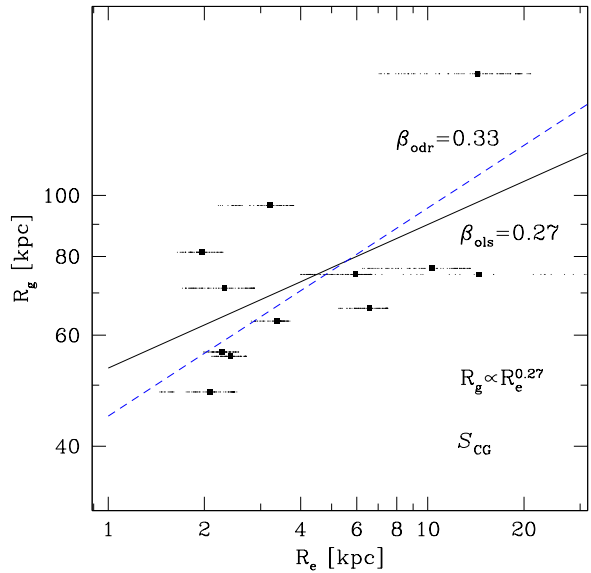}
\vspace{-0.8cm}
\caption{Similar as to Figure~2, but for the gravitational radius $R_{\rm g}$
and the effective radius $R_{\rm e}$. }
\label{fig:rads}
\end{figure}

In Figure~\ref{fig:rads} we plot the gravitational radius, $R_{\rm g}$,
 of each merger remnant versus their corresponding 
effective radii, $R_{\rm e}$,  obtained from a  fit using their S\'ersic 
growth-curve with $\xi=100\,$pc, and considering all the randomly generated   
projections in the sky.

 A fit in log-space of the form
\begin{equation}
R_{\rm g}  \propto R_{\rm e}^\beta \,
\end{equation}
was done. 
In Table~3 we list the different values of 
$\beta$  found under the different fittings conditions. Errors in $\beta$,
under both the {\sc ols} and {\sc odr}, are $\pm 0.01$.

The $\beta$ values do not change ($\approx 20$\%) as much as
when the fit is done with an  {\sc ols} or an {\sc odr}, in comparison to the
$\approx 50$\% change of the $\alpha$ index.
An increase in $\beta$ of $\approx 20$\% is obtained when 
the fitting is done on the profile
using a $\xi=1\,$kpc instead of $100\,$pc. We note that when the growth-curve
method is used the $\beta$ values decrease.

The $\beta$ index strongly depends on the adopted form of the fitting
law, irrespectively of the 
fitting procedure. Its variation can be as much as $\approx 200$\%
when a S\'ersic law is used instead of the $R^{1/4}$ law.
This behavior is most probably related to the one observed in observational
studies (e.g., Khosroshashi et~al.~2004), where totally
different values of $R_{\rm e}$ can be obtained if the brightness 
profile is not fitted by a suitable model. Here, the election of the
adopted profile is
reflected on the values of the $\beta$ index.

\begin{table}\label{tab:radius}
\begin{center}
\centering
\caption{Radial scalings $R_{\rm g}\propto R_{\rm e}^\beta$}
\begin{tabular}{|r|c|c|c|c|c|} \hline
$\xi$  & $R^{1/4}_{\rm P}$ & $R^{1/4}_{\rm{CG}}$ & $S_{\rm P}$ & $S_{\rm{CG}}$
 & {\sc FIT}    \\ \hline \hline
100 pc &  0.09 & 0.16 & 0.27 & 0.27 & {\sc ols} \\
       &  0.11 & 0.19 & 0.33 & 0.33 & {\sc odr} \\ \hline
1000 pc & 0.12 & 0.11 & 0.23 & 0.23 & {\sc ols} \\
        & 0.13 & 0.12 & 0.26 & 0.25 & {\sc odr}  \\  \hline
\end{tabular}
\end{center}
\end{table}

Averaging all the values listed in Table~3 related to the de Vaucouleurs law, 
irrespective of the fitting procedure or $\xi$ value, we obtain a mean of
$\langle \beta \rangle_{\rm{deV}}=0.13$, while for the S\'ersic law a value of
 $\langle \beta \rangle_{\rm{S}}=0.27$ is obtained. 

The dispersion of values, due to projection effects, around the 
mean value can be appreciated in Figure~\ref{fig:rads}; where 
a S\'ersic  growth-curve method is used to determine 
$R_{\rm e}$. Although not shown, the dispersion around the mean value is 
somewhat larger when a $R^{1/4}$ profile is fitted.

The results obtained for the $\beta$ index, under all the fitting conditions
considered, indicate a strong breaking of an homologous scaling 
between $R_{\rm g}$
and $R_{\rm e}$. In average, the $\beta$ index deviates $\approx 80$\% away 
from  the homology value of $\beta=1$.

\subsection{Masses}

\begin{figure}
\centering
\includegraphics[width=8cm]{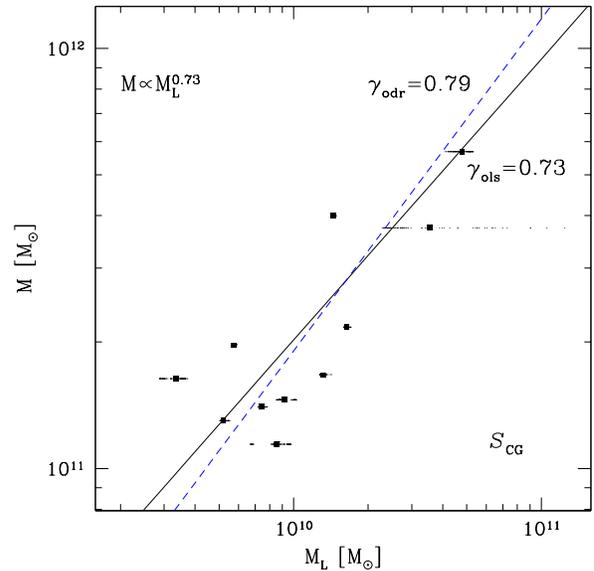}
\vspace{-0.8cm}
\caption{Similar to Figure~2, but for the total mass $M$ and the luminous mass
$M_{\rm L}$ inferred from the fit.}
\label{fig:mass}
\end{figure}

In Figure~\ref{fig:mass} we  plot the total mass of the remnant $M$
versus its luminous mass $M_{\rm L}$, where the latter was obtained using
a fit to a S\'ersic growth-curve and $\xi=100\,$pc. 

A relation of the form
\begin{equation}
M \propto M_{\rm L}^\gamma 
\end{equation}
is assumed 
for the fitting in log-space of the data set. 
In Table~4 we list the  values of the mass scaling index
$\gamma$ for all the different fittings considered here. Errors in $\gamma$
in both procedures are $\pm 0.01$.

The  $\gamma$ index shows a smaller dispersion against 
projection effects in comparison to the $\alpha$ or $\beta$ indexes. 
This is probably related
to the fact that $M_{\rm L}$ is an integrated quantity.

The mass index $\gamma$ shows the same tendency to increase its value when an
{\sc odr} fitting procedure is used in comparison to the {\sc ols} one.
Increasing the $\xi$ value has the effect of lowering the value of $\gamma$
 in all cases considered, and is rather stable under the two laws of
luminous matter distribution assumed (de Vaucouleurs or S\'ersic). 
Averaging all values under the {\sc odr} and
{\sc ols} procedures 
we find a $\langle \gamma \rangle_{\rm{odr}}=0.76$ and 
 a $\langle \gamma \rangle_{\rm{ols}}=0.69$, respectively.

It is difficult to adequately transform our $M_{\rm L}$ to a luminosity $L$ in 
order to compare our results with observations, this
 due to the uncertainties
on the mass-to-light ratios of ellipticals and the dissipationless nature
of our simulations.
Nonetheless, if we consider that $M \propto M_{\rm L}^{0.8}$, as provided by
an {\sc odr} fitting procedure using a S\'ersic law with $\xi=100\,$pc,  
then the following total mass-to-luminous mass ratio is obtained:
$$
\frac{M}{M_{\rm L}} \propto \frac{M_{\rm L}^{0.8}}{M_{\rm L}} \propto
M_{\rm L}^{-0.2} \;.
$$

\begin{table}\label{tab:mass}
\begin{center}
\centering
\caption{Mass scalings $M\propto M_{\rm L}^\gamma$}
\begin{tabular}{|r|c|c|c|c|c|} \hline
$\xi$  & $R^{1/4}_{\rm P}$ & $R^{1/4}_{\rm{CG}}$ & $S_{\rm P}$ & $S_{\rm{CG}}$
 & {\sc FIT}    \\ \hline \hline
100 pc  & 0.69 & 0.76 & 0.72 & 0.73 & {\sc ols} \\
        & 0.78 & 0.84 & 0.79 & 0.79 & {\sc odr} \\ \hline
1000 pc & 0.64 & 0.62 & 0.68 & 0.67 & {\sc ols} \\
        & 0.72 & 0.70 & 0.74 & 0.72 & {\sc odr}  \\  \hline
\end{tabular}
\end{center}
\end{table}

Assuming a constant mass-to-light ratio, the previous result would imply that 
$M/L \propto L^{-0.2}$. Observationally it has been found,
 using a S\'ersic profile (Trujillo, Burkert \& Bell 2004), 
that $M/L\propto L^{0.06\pm 0.04}$. 
When comparing this result with
the one obtained for our merger remnants, it follows that a constant
 mass-to-light ratio 
is not appropriate  to reproduce the observational results.

Nevertheless,  a constant mass-to-light ratio
 does not seem so unrealistic especially when mass-to-light ratios 
$\propto L^{-0.4}$ have been found for dwarf ellipticals
 (Peterson \& Caldwell 1993). Such dependency is more  closely reproduced, 
under a constant $M/L$ ratio, using other values of Table~4. 
However, a consistent comparison with observations requires additional physics 
not considered in this study.

\section{Discussion and Conclusions}

We have carried out twelve $N$-body simulations of binary mergers of
disk galaxies, constructed using the galaxy formation model of MMW and
with  properties consistent with a Tully-Fisher realization at $z\!=\!1$. 
For the merger remnants, scaling relations among the physical quantities 
that appear in the virial theorem and their observationally related ones 
were obtained.

In particular we looked for relations of the form: 
$V \propto \sigma_0^\alpha$, $R_{\rm g} \propto R_{\rm e}^\beta$, and
$M \propto M_{\rm L}^\gamma$. It is found that the scaling indexes 
 are sensitive to the fitting procedure ({\sc odr} or {\sc ols}), 
to the inner starting radius of the 
fitting region, $\xi$,  to the kind of law assumed to follow
the luminous matter (de Vaucouleurs or S\'ersic), and to whether 
 a profile or growth-curve is used.  
The  $\gamma$ index results to be the more stable under all
 these different fitting conditions.

In general, our results show that a strong breaking of homology
occurs in dissipationless mergers. 
We find that the $\alpha$ and 
$\gamma$ indexes are  more sensitive to the fitting procedure, obtaining
for the {\sc ols} procedure a $\langle \alpha \rangle_{\rm{ols}}=1.51$
  and  a 
$\langle \gamma \rangle_{\rm{ols}}=0.69$,  while for  the {\sc odr} procedure
  $\langle \alpha \rangle_{\rm{odr}}=2.35$ and 
  $\langle \gamma \rangle_{\rm{odr}}=0.76$.
The $\beta$ index results to be more sensitive on the assumed 
law for luminous matter distribution, values 
$\langle \beta \rangle_{\rm{deV}}=0.13$ and 
$\langle \beta \rangle_{\rm{S}}=0.27$ are found.

An immediate consequence of our results is the existence of a
 non-linear scaling
between the virial theorem, $M\propto V^2 R_{\rm g}$, and its observational
analogy, $M_{\rm L} \propto \sigma_0^2 R_{\rm e}$. Moreover, 
the ``constant''
 of proportionality between the physical and observational virial
relations depends on the fitting procedure and the radial range where the 
fit is done. 
This indicates that care has to be taken when trying to obtain
physical information from these observational parameters; for example, in 
determining the mass-to-light ratio of ellipticals using a kinematical 
approach based on the observational virial relation (e.~g., 
Padmanabhan~et al.~2004). Our results suggest
  that a dynamical mass estimate  
based on an homologous relation of the form 
$M_{\rm dyn} \propto \sigma_0^2 R_{\rm e}$ is likely to be incorrect.

Properties of remnants depend on the angular momentum and energy 
of the orbit of the progenitors
 (e.g., Naab \& Burkert 2003, Gonz\'alez-Garc\'{\i}a
 \& Balcells 2005, Boylan-Kolchin, Ma \& Quataert 2005). We expect that the
scaling indexes, that reflect the breaking of homology in remnants,  will also
depend on these quantities. Boylan-Kolchin et~al.~(2005) address 
 in an approximate manner, and with $N$-body simulations,
 the degree of homology breaking and find a 
dependence on the type of orbit considered. Although they
 just used one pericentric distance and a radial orbit, aside of using 
 only spherical models, their result 
 hints toward
a dependency on angular
momentum. Considering the small parameter 
space of simulations sampled out here, the values of 
$\{\alpha,\beta,\gamma\}$ obtained in this study are
 to be taken as indicative of 
the values they
might attain under more general conditions. 
A complete study will require to study
the dependence of the scaling indexes on the energy and angular momentum
of the orbit, and needs to be addressed in the future.

Comparing our $N$-body results directly to observations 
is restricted by the uncertainties in transforming the luminous mass, 
$M_{\rm L}$,
to luminosities, $L$. This is not an easy problem and would require  
to include, among other things, gas and stellar populations evolution
 models in the simulations. 
It is not clear at this stage how the scaling indexes would be 
affected by the inclusion of this new physics into the problem. 
However, in conclusion, it is clear from a purely $N$-body point
of view that homology is not satisfied in merger remnants of spiral galaxies.

\section*{Acknowledgments}

This research was funded by CONACyT-M\'exico Project 37506-E. An anonymous
referee is thanked for important comments that helped to improve the
presentation and content of this work.

\section*{APPENDIX A}

\subsection*{Orthogonal Fit to a Line}

 In the ordinary least-squares ({\sc ols}) method
 errors in the ``independent'' variable are minimized. However, if there is 
no  clear distinction which variable is
 dependent or independent a natural choice is to minimize errors in 
the normal direction to the surface fitted. 
This is the idea behind the orthogonal distance regression fitting
 ({\sc odr}). We describe here the procedure
 used to fit a line by this method.

\begin{figure}
\centering
\includegraphics[width=8cm]{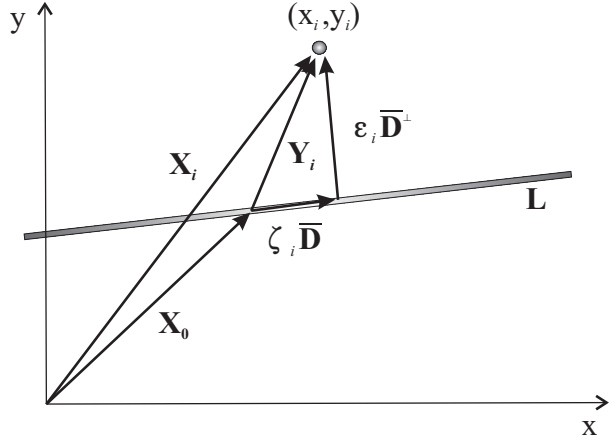}
\caption{Schematic diagram indicating the quantities used for fitting a line 
to a set of data points by orthogonal distance regression ({\sc odr}).}
\label{fig:lsqline}
\end{figure}

Let the line equation be described by
\begin{equation}
{\bf L} = {\bf X}_0 + t {\bf {\bar D}} \;,
\label{eq:line}
\end{equation}
where ${\bf {\bar D}}$ is a unit length vector along the line and $t$
some scalar. If 
${\bf X}_i$ is the 2-dimensional vector of the $n$ data points 
to be fitted, it can be written as
\begin{equation}
{\bf X}_i  = {\bf X}_0 + \zeta_i {\bf {\bar D}} + \varepsilon_i 
{\bf {\bar D}}^\perp \,,
\end{equation}
where ${\bf {\bar D}}^\perp$ is a unit vector perpendicular to 
${\bf{\bar D}}$ and $\zeta_i$, $\varepsilon_i$ 
are some scalars; see Figure~\ref{fig:lsqline}.  

Let  ${\bf Y}_i= {\bf X}_i - {\bf X}_0$, then we have that
$$
{\bf Y}_i - \zeta_i {\bf {\bar D}} = \varepsilon_i  {\bf {\bar D}}^\perp \;.
$$
The function to be minimized in  {\sc odr} is 
\begin{equation}
E({\bf X}_0,{\bf {\bar D}}) = \sum_{i=1}^n \varepsilon_i^2 = \sum_{i=1}^n
({\bf Y}_i - \zeta_i {\bf {\bar D}})^2 \;.
\end{equation}
We can write this as
\begin{equation}
E({\bf X}_0,{\bf {\bar D}}) = \sum_{i=1}^n [   {\bf Y}_i^T ( {\bf 1} - 
 {\bf {\bar D}} {\bf {\bar D}}^T )   {\bf Y}_i ] \,.
\label{eq:line1}
\end{equation}
The best fit ${\bf X}_0$ is obtained by setting
$$
\frac{\partial E}{\partial {\bf X}_0}= -2 ( {\bf 1} -  {\bf {\bar D}} 
{\bf {\bar D}}^T ) 
 \sum_{i=1}^n  {\bf Y}_i = 0 \,,
$$
that is obtained by setting $ \sum_{i=1}^n  ( {\bf X}_i - {\bf X}_0) = 0$, 
hence
\begin{equation}
{\bf X}_0 = \frac{1}{n} \sum_{i=1}^n   {\bf X}_i = 
\left[
\begin{array}{c}
\langle x \rangle  \\
\langle y \rangle  \\
\end{array}
\right] \,.
\label{eq:planeX0line}
\end{equation}

To obtain the best fit  ${\bf {\bf D}}$ 
we find de minimum of the following equivalent expression to (\ref{eq:line1}) 
$$
E({\bf X}_0,{\bf {\bar D}}) = {\bf {\bar D}}^T \left\{ 
 \sum_{i=1}^n  [ (  {\bf Y}_i \cdot {\bf Y}_i ) {\bf 1} - 
{\bf Y}_i {\bf Y}_i^T ] \right\}  {\bf {\bar D}} \;,
$$
and ${\bf 1}$ is the unit matrix. The term in brackets is a matrix, so we write this expression as
\begin{equation}
E({\bf X}_0,{\bf {\bar D}}) \equiv  
    {\bf {\bar D}}^T M({\bf X}_0) {\bf {\bar D}}\;.
\label{eq:planeM2}
\end{equation}
Writing explicitly matrix $M({\bf X}_0)$ we have
\begin{equation}
M({\bf X}_0) =
\left[
\begin{array}{cc}
 \delta_{yy}^2 & -\delta_{xy}^2  \\
 -\delta_{xy}^2 & \delta_{xx}^2 \\
\end{array}
\right] \,,
\label{eq:matrixM2}
\end{equation}
where  the $\delta_{xy}$ element is given by
\begin{equation}
\delta_{xy}^2=\sum_{i=1}^n \left( x_i - \langle x \rangle \right) 
 \left( y_i - \langle y \rangle \right) \,.
\end{equation}

From linear algebra we recognize the expression in (\ref{eq:planeM2}) as 
a quadratic form. Its minimum value 
is provided by the eigenvector corresponding to the lowest eigenvalue
 of matrix (\ref{eq:matrixM2}). In particular, normalizing 
this eigenvector to unity leads us directly the components 
 ${\bf {\bar D}}=(p,q)$, and hence the best fitted line (\ref{eq:line}). 
The line equation in parametric form can be written as
$$
x = x_0 + t p \;, \qquad
y = y_0 + t q \,,
$$
and, eliminating the parameter $t$, we have the best
 line fitted by {\sc odr} as:
\begin{equation}
y = \frac{q}{p} x + (y_0 - \frac{q}{p} x_0 ) \;.
\end{equation}
Eigenvalues may be found by using Jacobi method (Press et al. 1992).



\end{document}